\documentclass[conference, 9pt]{IEEEtran}
\usepackage{cite}
\usepackage{amsmath,amssymb,amsfonts}
\usepackage{algorithmic}
\usepackage{graphicx}
\usepackage{textcomp}
\usepackage{xcolor}

\usepackage{subcaption}     
\usepackage{makecell}       
\usepackage{multirow}       
\usepackage{color, colortbl}
\usepackage{pifont}         
\usepackage{enumitem}		
\usepackage{glossaries}     
\usepackage{xfrac}          
\usepackage{url}            
\usepackage{bibunits}
\usepackage{microtype}

\newif\iftr     
\trtrue

\newif\ifconf   
\conffalse

\newif\ifblind  
\blindfalse

\newif\ifcom    
\comfalse

\newif\ifps     
\psfalse

\setacronymstyle{long-short}
\newacronym{mcms}{MCMs}{multi-chip modules}
\newacronym{scps}{SCPs}{single-chip packages}
\newacronym{d2d}{D2D}{die-to-die}
\newacronym{usr}{USR}{ultra-short reach}
\newacronym{c4}{C4}{controlled collapse chip connection}
\newacronym{phy}{PHY}{physical layer}
\newacronym{g}{\texttt{G}}{grid}
\newacronym{hc}{\texttt{HC}}{honeycomb}
\newacronym{bw}{\texttt{BW}}{brickwall}
\newacronym{cor}{\texttt{HM}}{HexaMesh}
\newacronym{ici}{ICI}{inter-chiplet interconnect}
\newacronym{pcb}{PCB}{printed circuit board}

\def\BibTeX{{\rm B\kern-.05em{\sc i\kern-.025em b}\kern-.08em
    T\kern-.1667em\lower.7ex\hbox{E}\kern-.125emX}}

\definecolor{lightgray}{gray}{0.8}

\ifcom
\usepackage{color-edits}    
\addauthor[Summary]{ps}{orange}
\addauthor[Patrick]{pi}{teal}
\addauthor[Maciej]{mb}{blue}
\addauthor[Torsten]{th}{red}
\addauthor[Matheus]{mc}{orange}
\addauthor[Tim]{tf}{violet}
\addauthor[L. Benini]{lb}{violet}
\fi

\newcommand{\picom}[1]{\ifcom\picomment{#1}\fi}
\newcommand{\mbcom}[1]{\ifcom\mbcomment{#1}\fi}

\newcommand{\lbcom}[1]{\ifcom\lbcomment{#1}\fi}

\newcommand{\ps}[1]{\ifps\pscomment{#1}\fi}

\setlist{leftmargin=1em}

\begin{document}

\title{
HexaMesh: Scaling to Hundreds of Chiplets\\with an Optimized Chiplet Arrangement
\ifconf\vspace{-2em}\fi
}

\ifblind
\author{}
\else
\author{
\IEEEauthorblockN{
Patrick Iff\IEEEauthorrefmark{1}, 
Maciej Besta\IEEEauthorrefmark{1}, 
Matheus Cavalcante\IEEEauthorrefmark{2},
Tim Fischer\IEEEauthorrefmark{2}, 
Luca Benini\IEEEauthorrefmark{2}\IEEEauthorrefmark{3} and
Torsten Hoefler\IEEEauthorrefmark{1}
}
\IEEEauthorblockA{
\IEEEauthorrefmark{1}Department of Computer Science,
ETH Zurich, Zurich, Switzerland\\
Email: \{patrick.iff, maciej.besta, htor\}@inf.ethz.ch}
\IEEEauthorblockA{
\IEEEauthorrefmark{2}Department of Information Technology and Electrical Engineering,
ETH Zurich, Zurich, Switzerland\\
Email: \{matheus, fischeti, lbenini\}@iis.ee.ethz.ch}
\IEEEauthorblockA{
\IEEEauthorrefmark{3}
Dept. of Electrical, Electronic and Information Engineering, University of Bologna, Italy}
}
\fi

\maketitle
\begin{abstract}
2.5D integration is an important technique to tackle the growing cost of manufacturing chips in advanced technology nodes.
This poses the challenge of providing high-performance inter-chiplet interconnects (ICIs).
As the number of chiplets grows to tens or hundreds, it becomes infeasible to hand-optimize their arrangement in a way that maximizes the ICI performance.
In this paper, we propose HexaMesh, an arrangement of chiplets that outperforms a grid arrangement both in
theory (network diameter reduced by $42\%$; bisection bandwidth improved by $130\%$) 
and in practice (latency reduced by $19\%$; throughput improved by $34\%$).
HexaMesh enables large-scale chiplet designs with high-performance ICIs.

\end{abstract}


\section{Introduction}
\label{sec:intro}

\ps{Scaling challenges: Economics}

CMOS technology scaling enables us to build chips with an ever-increasing transistor density.
The main advantage of transitioning to a more advanced technology node is that it allows 
us to pack more transistors and hence more performance into chips of the same size.
The downside of this transition is the increased complexity of the physical design, 
verification, firmware, and mask sets.
As a result, the non-recurring cost almost doubles whenever we transition to a more advanced 
technology node \cite{design-cost-exp}.
Another challenge of manufacturing chips in advanced technology nodes is the high
defect rate which diminishes the yield and increases the recurring cost.
Due to these trends, making the design and fabrication of chips in bleeding-edge
technology nodes economically viable has become a real challenge.

\ps{Chiplet motivation: Solve scaling challenges}

A promising solution to this challenge is the disaggregation of monolithic chips into
\gls{mcms}. Current trends show that the only chips 
that keep up with Moore's law \cite{mooreslaw} are \gls{mcms}~\cite{lookingglass2022}. 
One strategy to create multi-chip modules is 2.5D integration in which the chip is 
disaggregated into multiple chiplets which are connected through an organic packaging 
substrate or silicon interposer.
2.5D integration has various economical advantages:
\lbcom{if you need space, this itemize can be reduced in size and fit in a single paragraph. 
The DAC audience and reviewers do not need to be tutored on these advantages of chiplets!}
\picom{I tried to make the itemize more compact by reducing the word-count.}
\begin{itemize}
	\item	\textbf{Heterogeneity}: Different chiplets can be implemented in different 
			technology nodes. Here, subcircuits that cannot take advantage
			of transistor scaling, e.g., I/O drivers, are fabricated in 
			more mature technology nodes with lower non-recurring cost and higher yield.
	\item	\textbf{Reuse}: A given chiplet can be used in multiple designs. For example,
			we do not need to redesign the aforementioned I/O chiplet when the rest of
			the chip is transitioned to a more advanced technology node. As demonstrated by 
			AMD \cite{amd-chiplets}, we can use the same compute-chiplet
			in multiple products with varying core-counts. Reuse avoids
			redesigning components, further reducing the non-recurring cost.
	\item	\textbf{Improved Yield}: A single fabrication defect can render a whole die 
			useless, whether it is a chiplet or a monolithic chip. Since chiplets are smaller
			than monolithic chips, 2.5D integration reduces the area loss due to 
			fabrication defects, hence improving the yield.
	\item	\textbf{Binning}: Power- and frequency-binning are important strategies to 
			deal with parametric variation. In binning, chips are grouped into different
			bins (e.g., based on power consumption or maximum clock frequency) which are
			then priced differently. In 2.5D integration, binning is done on a 
			per-chiplet scale, increasing the total revenue.
\end{itemize}

\ps{Chiplet challenges: Interconnect performance}

While 2.5D integration has many economical benefits, it also comes with technological 
challenges. One such challenge is the fact that a \gls{d2d} link requires a \gls{phy} interface in both 
the sending and the receiving chiplet. As a consequence, the total silicon area and power 
consumption of all chiplets combined exceed the area and power of a monolithic chip with the
same functionality. However, the additional cost due to the \gls{phy}'s area and power overhead is 
often compensated by the other economical benefits of 2.5D integration. 
A more important challenge is creating a high-bandwidth and low-latency \gls{ici}.
To connect chiplets to the package substrate, \gls{c4} bumps are used and to connect
them to a silicon interposer, one uses micro-bumps.
The minimum pitch of these bumps limits the number of bumps per 
mm$^2$ of chiplet area which limits the number and bandwidth of \gls{d2d} links.
As a consequence, \gls{d2d} links are the bottleneck of the \gls{ici}.

\ps{Why shapes and arrangements matter}

Since the \gls{d2d} links limit the \gls{ici} data width, we want to operate them at 
the highest frequency possible to maximize
their throughput. To run such links at high frequencies without introducing unacceptable
bit error rates, we must limit their length to a minimum \cite{bow, ucie}.
The length of \gls{d2d} links is minimized if we only connect adjacent chiplets.
However, with such restricted connections, the shape and 
arrangement of chiplets has a significant impact on the
performance of the \gls{ici}.

\ps{Our contributions}

In this paper, we analyze how to shape and arrange chiplets to maximize the 
\gls {ici} performance. We make the following contributions:
\begin{itemize}
    \item   \textbf{Problem Statement}:
            We formulate a detailed problem statement including economics- and technology-driven
            constraints for the shape of chiplets and proxies for the \gls{ici} performance
            (Section \ref{sec:problem}).
    \item   \textbf{HexaMesh}:
            We address the above problem by proposing the HexaMesh arrangement.
            HexaMesh asymptotically reduces the network diameter by $42\%$ and improves 
            the bisection bandwidth by $130\%$ compared to a grid arrangement
            (Section \ref{sec:proposal}).
    \item   \textbf{\gls{d2d} link model}:
            Instead of only relying on network diameter and bisection bandwidth as performance proxies, we
            want to consider implementation details of the \gls{ici}.
            To do so, we introduce our model to estimate the bandwidth of \gls{d2d} links
            (Section \ref{sec:model}).
    \item   \textbf{Evaluation}:
            We combine link bandwidth estimates using our model and cycle-level simulations
            using BookSim2 \cite{booksim} to compare HexaMesh to a 2D grid.
            On average, HexaMesh reduces the latency by $19\%$
            and improves the throughput by $34\%$ 
            (Section \ref{sec:evaluation}).
\end{itemize}

\section{Background on 2.5D Integration}
\label{sec:back}

\ps{Argue why we focus on (passive) silicon interposer and organic package substrate}

\begin{figure}[h]
\centering
\captionsetup{justification=centering}
\begin{subfigure}{0.98 \columnwidth}
\centering
\includegraphics[width=1.0\columnwidth]{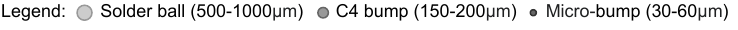}
\end{subfigure}
\begin{subfigure}{0.32 \columnwidth}
\centering
\includegraphics[width=1.0\columnwidth]{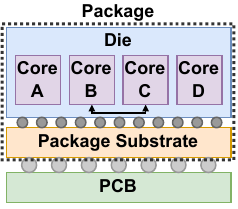}
\caption{Monolithic chip.}
\label{fig:back-integration-monolithic}
\end{subfigure}
\begin{subfigure}{0.32 \columnwidth}
\centering
\includegraphics[width=1.0\columnwidth]{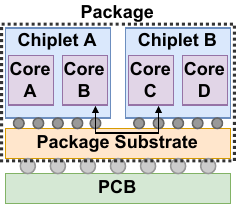}
\caption{Package substrate.}
\label{fig:back-integration-substrate}
\end{subfigure}
\begin{subfigure}{0.32 \columnwidth}
\centering
\includegraphics[width=1.0\columnwidth]{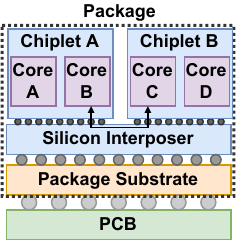}
\caption{Silicon interposer.}
\label{fig:back-integration-interposer}
\end{subfigure}
\caption{Comparison of a monolithic chip and 2.5D stacked chips using a package substrate or silicon interposer (side view).}
\label{fig:back-integration}
\vspace{-1em}
\end{figure}

In 2.5D integration, multiple chiplets are enclosed in a single package.
The two most prominent techniques to provide connectivity between chiplets 
are organic package substrates (see Figure \ref{fig:back-integration-substrate}) and
 silicon interposers (see Figure \ref{fig:back-integration-interposer}). 
Besides these established 2.5D integration schemes, there are more complex 
techniques, e.g., active interposers \cite{intact}. Active interposers do not only contain 
wires but also transistors which allows constructing buffered wires or offloading some 
power management circuits from the chiplets to the interposer. However, active interposers
come with additional challenges, e.g., reduced yield or thermal 
problems. In this work, we focus on passive silicon interposers and package substrates as
they are more established.

\ps{Explain organic package substrate}

The \textbf{organic package substrate} provides connectivity between different
chiplets and between chiplets and the \gls{pcb}. 
\gls{c4} bumps with a pitch of $150$-$200\mu$m are used to
connect chiplets to the package substrate. 
Connections between the package substrate and the \gls{pcb} are built using solder bumps with  
a pitch of $500$-$1000\mu$m. The small pitch of \gls{c4} bumps enables the construction of 
\gls{d2d} links that offer up to $44\times$ more bandwidth than off-chip links.
This shows that the bandwidth between multiple chiplets in a 2.5D stacked chip is substantially higher
than the bandwidth between multiple \gls{scps} on the same \gls{pcb}.

\ps{Explain pros and cons of silicon interposer}

A \textbf{silicon interposer} can be added between the chiplets and the package substrate. 
Micro-bumps with a pitch of $30$-$60\mu$m are used to connect chiplets to the 
interposer. Regular \gls{c4} bumps with a pitch of $150$-$200\mu$m are used to connect the 
interposer to the package substrate. The reduced pitch of micro-bumps further enhances the 
throughput of \gls{d2d} links. Besides increased design and manufacturing cost, silicon 
interposers also come with higher signal loss compared to package substrates \cite{usr-links}. 
As a consequence, \gls{d2d} links in silicon interposers need to be even shorter ($\leq 2$mm \cite{ucie})
to provide low bit error rates when operated at high frequencies.

\ps{Explain PHYs}

\textbf{\gls{d2d} links} often use different protocols, voltage levels, and clock frequencies 
than the intra-chiplet interconnect.
The conversion between protocols, voltage levels, and clock frequencies is performed by
a \gls{phy} which is added at the start and end of each \gls{d2d} link.
\gls{phy}s reside inside the chiplets and they introduce a certain area and power overhead compared to
monolithic chips that do not require them. \gls{phy}s and \gls{ici} protocols have been standardized 
\cite{bow, ucie} to achieve interoperability between chiplets from different manufacturers.

\section{The Problem of Chiplet Shape \& Arrangement}
\label{sec:problem}

\ps{Section intro}

In this section, we formalize the problem of finding chiplet shapes and arrangements. To do 
so, we define technology- and economics-driven constraints for the shape of chiplets.
We also introduce proxies for the performance of the \gls{ici} be able to
assess a given arrangement without making any assumptions on implementation details.

\subsection{Assumptions and Scope}
\label{ssec:problem-assumptions}

\ps{We search arrangements for N identical chiplets}

\begin{figure}[h]
\centering
\captionsetup{justification=centering}
\includegraphics[width=0.70\columnwidth]{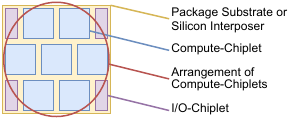}
\caption{We place chiplets for I/O drivers or other functions on the perimeter of
a proposed arrangement of compute-chiplets (top view).}
\label{fig:problem-assumptions}
\vspace{-1em}
\end{figure}

We assume that our chip consists of several identical compute-chiplets 
and additional chiplets for I/O drivers or other functions. We limit our scope to 
the search for shape and arrangement of the identical compute-chiplets. Whenever we propose
a shape and arrangement of compute-chiplets, we implicitly assume that the remaining chiplets 
are placed on the perimeter of our arrangement (see Figure \ref{fig:problem-assumptions}).
Placing the I/O drivers close to the border of the chip 
is favorable because usually, only solder balls at the border of the package are used for 
signals. As it is hard to route \gls{pcb} lanes to solder balls at the center of the package, 
those solder balls are often used for the power supply.

\subsection{Constraints for Chiplet Shapes}
\label{ssec:problem-constraints}

\ps{Chiplets must be identical and rectangular}

To ensure that we only consider designs that are economical and easy to manufacture, 
we identify constraints for the shape of chiplets.
\begin{itemize}
	\item	\textbf{Uniform Chiplets}: All compute-chiplets in a given arrangement must have the same shape and 
			size. Integrating the same functionality into multiple chiplets with different 
			shapes is technologically feasible, however, designing multiple compute-chiplets 
			for a single product generation would increase the non-recurring cost and 
			diminish the economical advantages of 2.5D integration.
	\item	\textbf{Rectangular Chiplets}: All chiplets must be rectangular. Dicing methods
			such as stealth dicing \cite{stealth-dicing} or plasma dicing \cite{plasma-dicing}
			enable the fabrication of non-rectangular chiplets. However, the most common dicing
			method is blade dicing which can only produce rectangular chiplets.
			By limiting our search to rectangular chiplets, we ensure that we only
			consider designs with a wide range of applicability.
\end{itemize}

\subsection{Proxies for Inter-Chiplet Interconnect Performance}
\label{ssec:problem-proxies}

\ps{Why performance proxies?}

How a given arrangement of chiplets translates into performance (latency and throughput) 
of the \gls{ici} is not obvious. We could run simulations, but this would force us to 
make many assumptions, e.g., on the bump pitch, chiplet area, or \gls{ici} communication protocol details.
To be able to predict the performance of an arrangement of chiplets without making any
assumptions, we introduce performance proxies. 

\ps{Introduce graph representation of 2.5D stacked chipls}

\begin{figure}[h]
\centering
\captionsetup{justification=centering}
\begin{subfigure}{0.6 \columnwidth}
\centering
\includegraphics[width=1.0\columnwidth]{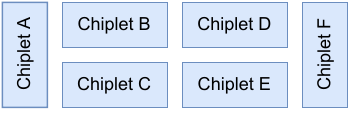}
\caption{Arrangement of chiplets (top view).}
\label{fig:probelm-example-chiplets}
\end{subfigure}
\begin{subfigure}{0.38 \columnwidth}
\centering
\includegraphics[width=0.9\columnwidth]{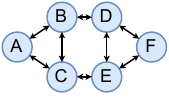}
\caption{Graph representation.}
\label{fig:probelm-example-graph}
\end{subfigure}
\caption{We represent arrangements of chiplets as graphs.}
\label{fig:problem-example}
\end{figure}

As discussed in
Section \ref{sec:intro}, we want to minimize the length of \gls{d2d} links which 
implies that only adjacent chiplets can be connected. To be more precise, we 
define that only chiplets sharing a common edge can be connected. We do not 
allow links between chiplets that only share a common corner as 
this would increase the link length.
Based on this definition, we represent our 2.5D stacked chip as a planar graph \cite{graphs} where vertices
correspond to chiplets and edges correspond to links. Two vertices are connected
by an edge whenever the corresponding chiplets are adjacent 
(see Figure \ref{fig:problem-example}).

\ps{Use network diameter and bisection bandwidth as proxies for latency and throughput}

We use a 2.5D stacked chip's graph representation to obtain proxies for the latency and global
throughput of the \gls{ici}.
Whenever a flit\footnote{Flow control unit: Atomic amount of data transported across the network.}
transitions from a chiplet to the interposer or vice-versa, it 
needs to be processed by a PHY which adds a certain latency. Based on this observation, we 
use the diameter of a chip's graph representation as a proxy for its latency
and we use the bisection bandwidth of the graph representation as a proxy for the global throughput. 

\subsection{Problem Statement}

\ps{Maximize proxies while fulfilling constraints}

In this work, we want to solve the following problem: 

\begin{center}
\textit{
Find a shape and arrangement of chiplets that maximizes the
proxies for inter-chiplet interconnect performance as defined in Section \ref{ssec:problem-proxies}
while satisfying all constraints from Section \ref{ssec:problem-constraints}.}
\end{center}

\begin{figure*}[h]
\centering
\captionsetup{justification=centering}
\begin{subfigure}[t]{0.22 \textwidth}
\centering
\includegraphics[width=0.75\columnwidth]{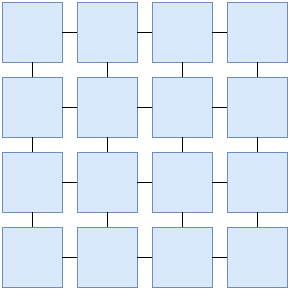}
\caption{Grid (\texttt{G}).\\Neighbors: Min: 2, Max: 4.\\Diameter: $2\sqrt{N}-2$.\\
Bisection BW: $\sqrt{N}$.}
\label{fig:proposal-arrangements-grid}
\end{subfigure}
\hspace{-1em}
\begin{subfigure}[t]{0.035 \textwidth}
\includegraphics[width=1.0\columnwidth]{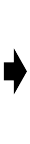}
\end{subfigure}
\hspace{+0em}
\begin{subfigure}[t]{0.22 \textwidth}
\centering
\includegraphics[width=1.0\columnwidth]{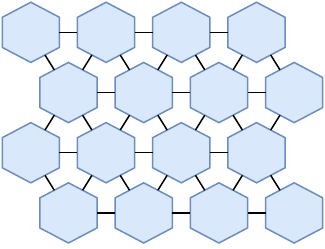}
\caption{Honeycomb (\texttt{HC}).\\
Neighbors: Min: 2, Max: 6.\\
Diameter: $2\sqrt{N}-2-\lfloor \frac{\sqrt{N}-1}{2}\rfloor$.\\
Bisection BW: $2\sqrt{N}-1$.}
\label{fig:proposal-arrangements-hex}
\end{subfigure}
\hspace{+0em}
\begin{subfigure}[t]{0.035 \textwidth}
\includegraphics[width=1.0\columnwidth]{img/proposal/arrow.drawio.pdf}
\end{subfigure}
\hspace{-1em}
\begin{subfigure}[t]{0.22 \textwidth}
\centering
\includegraphics[width=0.85\columnwidth]{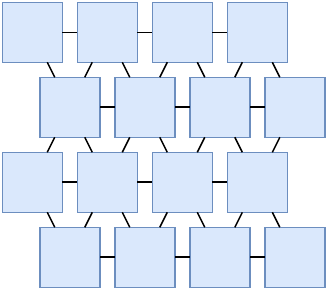}
\caption{Brickwall (\texttt{BW}).\\
Neighbors: Min: 2, Max: 6.\\
Diameter: $2\sqrt{N}-2-\lfloor \frac{\sqrt{N}-1}{2}\rfloor$.\\
Bisection BW: $2\sqrt{N}-1$.}
\label{fig:proposal-arrangements-brick}
\end{subfigure}
\hspace{-0.5em}
\begin{subfigure}[t]{0.035 \textwidth}
\includegraphics[width=1.0\columnwidth]{img/proposal/arrow.drawio.pdf}
\end{subfigure}
\hspace{-1.0em}
\begin{subfigure}[t]{0.22 \textwidth}
\centering
\includegraphics[width=0.75\columnwidth]{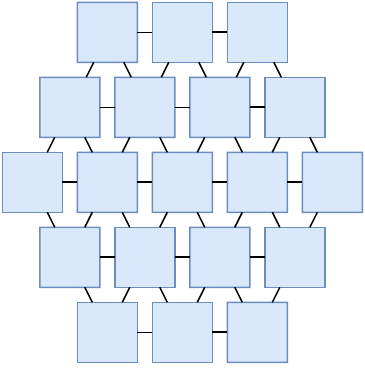}
\caption{HexaMesh (\texttt{HM}).\\
Neighbors: Min: 3, Max: 6.\\
Diameter: $\frac{1}{3}\sqrt{12N-3}-1$.\\
Bisection BW: $\frac{2}{3}\sqrt{12N-3}-1$.}
\label{fig:proposal-arrangements-circle}
\end{subfigure}
\caption{Evolution of compute-chiplet arrangements: From grid (the main baseline) to HexaMesh (our final design). 
Additional I/O chiplets are placed on the edge of those arrangements to fill the gap between non-rectangular arrangements and the rectangular package substrate or silicon interposer.
For each arrangement, we show the network diameter and bisection bandwidth as a function of the chiplet count $N$.}
\label{fig:proposal-arrangements}
\vspace{-1em}
\end{figure*}

\section{Enhancing Shape and Arrangement of Chiplets}
\label{sec:proposal}

\ps{Section intro}

We now illustrate a novel arrangement of chiplets, the \gls{cor}, which enhances the performance of
the \gls{ici} while maintaining ease of manufacturing.
For this, we first observe that
the most straightforward way to build a 2.5D stacked chip is arranging chiplets
in a 2D \gls{g}, which we use as the main baseline. 
We illustrate how to go through multiple improvements of a \gls{g} until we arrive at the \gls{cor}.
For each arrangement, we deliver a shape of chiplets and a placement of \gls{c4} bumps or 
micro-bumps that minimizes the length of \gls{d2d} links.
Finally, we discuss how to apply arrangements to arbitrary chiplet counts and we compare 
multiple arrangements in terms of their network diameter and bisection bandwidth
(performance proxies).

\subsection{Optimizing the Arrangement of Chiplets}
\label{ssec:proposal-arrangement}

\ps{Introduce grid, give intuition that a higher average number of neighbors is desirable}

\paragraph{Grid (\texttt{G})} We show a 2D grid in Figure \ref{fig:proposal-arrangements-grid}.
We observe that each non-border chiplet is connected to four other chiplets. 
Mathematically speaking, the average number of neighbors per chiplet goes to four as the 
number of chiplets goes to infinity. 
Intuitively, increasing the average number of neighbors per chiplet should reduce the network
diameter and increase the bisection bandwidth. 

\ps{Argue why we discuss hexagonal chiplets}

To explore arrangements that maximize the average number of neighbors per chiplet, 
we drop the constraint that chiplets need to be rectangular for the next paragraph. 
In the subsequent paragraph, we show how to fix this violation of constraints.

\ps{Introduce honeycomb, show that they asymptotically maximize the avg \#neighbors}

\paragraph{Honeycomb (\texttt{HC})} 
If we manufacture hexagonal chiplets and arrange them in a honeycomb
pattern, then, each non-border chiplet is connected to six other chiplets
(see Figure \ref{fig:proposal-arrangements-hex}).
The average number of neighbors per chiplet approaches six as the number 
of chiplets goes to infinity.
As we have seen in Section \ref{ssec:problem-proxies}, we can represent each arrangement of
chiplets as a planar graph (a graph that can be drawn such that no edges cross each other).
A fundamental theorem of graph theory states that for planar graphs with $v \geq 3$ vertices 
and $e$ edges, $e \leq 3v-6$ does hold.
We use this inequality to derive an upper bound for the average vertex degree $d_\text{avg}$ in 
planar graphs, which corresponds to the average number of neighbors per chiplet:

\begin{equation*} \label{eq:proposal-bound-neighbors}
	d_\text{avg} = \frac{2e}{v} \leq \frac{2 (3v - 6)}{v} = 6 - \frac{12}{v}.
\end{equation*}

Asymptotically speaking, the \gls{hc} maximizes the average 
number of neighbors per chiplet. 
However, it does violate our constraints since it uses
non-rectangular chiplets.

\ps{Introduce brickwall, show that it has the same advantages as honeycomb but without 
violating any constraints.}

\paragraph{Brickwall (\texttt{BW})} Arranging rectangular chiplets in a brickwall pattern
(see Figure \ref{fig:proposal-arrangements-brick}) results in the same graph structure 
as the \gls{hc}. This enables an asymptotically optimal 
average number of neighbors per chiplet without violating any constraints on the shape of chiplets.

\ps{Introduce circular, motivation: increasing the minimum number of neighbors, reducing the 
diameter.}

\paragraph{HexaMesh (\texttt{HM})} We want to further optimize our arrangement of chiplets. 
One issue in the \gls{bw} is that there are two chiplets with only two neighbors.
By arranging chiplets in a circle around one central chiplet 
(see Figure \ref{fig:proposal-arrangements-circle}), we can increase the minimum
number of neighbors per chiplet from $2$ to $3$. An additional advantage of this
arrangement is that it asymptotically reduces the network diameter by $33\%$ 
compared to the \gls{bw} (see Section \ref{ssec:proposal-proxies} for details).

\ps{Argue why we do not consider the honeycomb arrangement for the remainder of the paper}

As the \gls{hc} violates our constraints and the \gls{bw} results in 
the same graph structure, we only consider the \gls{g}, \gls{bw}, and 
\gls{cor} from now on.

\subsection{Optimizing the Shape of Chiplets}
\label{ssec:proposal-shape}

\ps{Subsection Intro}

For each arrangement that we discussed above, we find a shape of chiplets that maximizes
the performance of the \gls{ici}.

\lbcom{Since you have quite a feq symbols, a table with the defs could be useful.}
\picom{Due to space constraints, I didn't add the table, but I added more symbols to figure 5 which hopefully explains them visually.}

\ps{Divide chiplet area into sectors based on bump usage}

\begin{figure}[h]
\vspace{-1em}
\centering
\captionsetup{justification=centering}
\begin{subfigure}{0.43 \columnwidth}
\centering
\includegraphics[width=1.0\columnwidth]{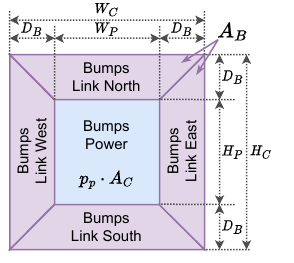}
\caption{Grid (\texttt{G}).~~~~}
\label{fig:proposal-bumps-grid}
\end{subfigure}
\begin{subfigure}{0.55 \columnwidth}
\centering
\includegraphics[width=0.83\columnwidth]{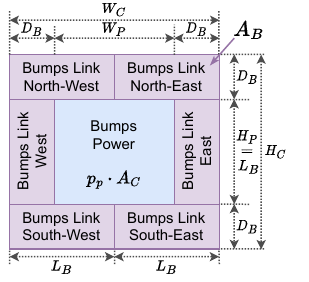}
\vspace{-1.0em}
\caption{Brickwall (\texttt{BW}) and HexaMesh (\texttt{HM}).}
\label{fig:proposal-bumps-brick}
\end{subfigure}
\caption{Assignment of \gls{c4} bumps or micro-bumps in chiplets.}
\label{fig:proposal-bumps}
\vspace{-1em}
\end{figure}

Recall that the \gls{ici} is
built using \gls{d2d} links which are attached to chiplets using \gls{c4} bumps or micro-bumps. The 
bandwidth of a link is larger if the link has more bumps at its disposal. The maximum number
of bumps per chiplet is proportional to the chiplet area $A_C$. A fraction $p_p \in [0,1]$ of 
these bumps is used for the chiplet's power supply and the remaining bumps are
used for \gls{d2d} links. We divide the area of a
chiplet into different sectors. Each sector contains bumps used for either the power 
supply or for one of the \gls{d2d} links (see Figure \ref{fig:proposal-bumps}).
To make sure that all links have the same bandwidth, all sectors for bumps
of \gls{d2d} links must have the same area $A_B$.

\ps{Minimize distance between bumps and chiplet edge}

The second shape-related factor that influences the performance of the \gls{ici} 
is the length of \gls{d2d} links. 
To minimize the link length, we minimize the maximum distance $D_B$ 
between a bump and the edge of the chiplet (see Figure \ref{fig:proposal-bumps}).
To minimize $D_B$, we place the sector containing power bumps in the 
center of the chiplet and we place the sectors for bumps of \gls{d2d} links at the chiplet edges. 
To make sure that all \gls{d2d} links have the same performance, we enforce that the distance
$D_B$ is identical for all sectors containing link bumps.

\ps{Bump assignment and chiplet shape for grid}

\paragraph{Grid (\texttt{G})}
Figure \ref{fig:proposal-bumps-grid} displays how we arrange \gls{c4} bumps or micro-bumps in chiplets
of the \gls{g}. The measurements annotated in said figure guarantee that
the maximum distance $D_B$ between a bump and the edge of the chiplet is identical for
all links.
\mbcom{this is repeating, same as above, merge/eliminate redundancy in text.}
\picom{In my opinion, this is not redundant. In the previous paragraph, we state why we NEED identical $D_B$ and $A_B$ (in general)
and in this paragraph, we explain how we GUARANTEE identical $D_B$ and $A_B$ in our proposed chiplet shape for the G.
And in the next paragraph we explain how we GUARANTEE identical $D_B$ and $A_B$ in our proposed chiplet shape for the BW and CoR.}
\lbcom{Tend to agree}
To guarantee that all sectors for link bumps have the same
area $A_B$, we require that the chiplets are square ($W_C = H_C = \sqrt{A_C}$) which implies 
that the sector for power-bumps is square ($W_P = H_P = \sqrt{p_p \cdot A_C}$). 
The area of one sector for bumps of a \gls{d2d} links is 
$A_B = (1/4) (1-p_p) A_C$ and the maximum distance between a bump and the edge
of the chiplet is $D_B = (W_C - W_P) / 2 = (H_C - H_P) / 2$.

\ps{Bump assignment and chiplet shape for brickwall and hexamesh}

\paragraph{Brickwall (\texttt{BW}) and HexaMesh (\texttt{HM})}
For the \gls{bw} and \gls{cor}, we arrange the \gls{c4} bumps or micro-bumps as
displayed in Figure \ref{fig:proposal-bumps-brick}. Similarly to the \gls{g}, the measurements annotated in said 
figure guarantee that both the area of each sector for link bumps $A_B$ and the maximum 
distance between a link bump and the edge of the chiplet $D_B$ are identical for all \gls{d2d} links.
The area available for bumps of a given link is $A_B = (1/6) (1-p_p) A_C$.
Computing the maximum distance $D_B$ between a link bump and the edge of the chiplet
as well as the resulting chiplet dimensions is a bit more involved.
Based on Figure \ref{fig:proposal-bumps-brick}, we set up the following system of equations:
\begin{equation} \label{eq:bumps-1}
	H_C = 2D_B + L_B
\end{equation}
\begin{equation} \label{eq:bumps-2}
	W_C = 2L_B
\end{equation}
\begin{equation} \label{eq:bumps-3}
	W_P = W_C - 2 D_B
\end{equation}
\begin{equation} \label{eq:bumps-4}
	H_C \cdot W_C = A_C
\end{equation}
\begin{equation} \label{eq:bumps-5}
	W_P \cdot L_B = A_C \cdot p_p
\end{equation}
By solving this system of equations, we get the chiplet dimensions $W_C$ and $H_C$ as well
as the maximum distance $D_B$ between a bump and the edge of the chiplet:
\begin{equation*} \label{eq:bumps-11}
	W_C = \sqrt{\frac{A_C ( 2 + 4 p_p)}{3}}
	\quad
	H_C = \frac{A_C}{W_C}
	\quad
	D_B = \frac{(1-p_p)A_C}{\sqrt{A_C(6 + 12p_p)}}
\end{equation*}

\ps{Example for equations above (suggested by TH)}

Consider an example design with a chiplet area of $A_C = 16$ mm$^2$ where a fraction 
$p_p = 0.4$ of all bumps are needed for the power supply. Our equations yield
the chiplet dimensions $W_C = 4.38$ mm and $H_C = 3.65$ mm and a
maximum distance of $D_B = 0.73$ mm between a bump used for \gls{d2d} links and the 
chiplet edge.

\subsection{Applicability of Arrangements}

\ps{Arrangements that only apply to certain chiplets counts are a problem}

To apply the \gls{g} or \gls{bw} as depicted 
in Figure \ref{fig:proposal-arrangements}, the number of chiplets $N$ needs to be a square 
number and for the \gls{cor}, we need to have $N = 1 + 3 r (r+1)$ for some 
$r \in \mathbb{N}$ (if there are $r$ rings around the central chiplet where the $i$-th 
ring contains $6i$ chiplets, then, we have $1 + \sum_{i=1}^r 6i = 1 + 3 r (r+1)$ chiplets in total).
We call such an arrangement
\textit{regular}. For the \gls{g} and \gls{bw}, we could also use $R$ rows and $C$
columns of chiplets such that $RC = N$, but $R \neq C$ which results in a rectangular, 
non-square shape. We call this a \textit{semi-regular} arrangement. Semi-regular
arrangements make only sense if $R$ and $C$ are similar, otherwise, both diameter and
bisection bandwidth deteriorate. We conclude that for many chiplet-counts, there is no
regular or no reasonable semi-regular arrangement. This is a problem as we want to set the 
number of chiplets based on technological and economical factors, not their desired arrangement.

\ps{Solve this problem by using irregular arrangements}

To solve this problem, we introduce \textit{irregular} arrangements.
Starting from the closest smaller regular arrangement, we incrementally add
more chiplets until the desired chiplet-count is reached. In the case of the
\gls{g} and \gls{bw}, these additional chiplets form 
incomplete rows or columns, and in the case of the \gls{cor},
they form an incomplete circle.

\ps{Disadvantages of irregular arrangements}

For regular and semi-regular \gls{g} and \gls{bw} of $N \geq 4$ chiplets, the 
minimum number of neighbors per chiplet is $2$, and for regular \gls{cor} of 
$N \geq 7$ chiplets, the minimum number of neighbors per chiplet is $3$. 
Introducing irregular arrangements reduces the minimum number of neighbors per chiplet
to $1$ for some \gls{g} and to $2$ for some \gls{cor}.
This suggests that irregular \gls{g} and \gls{cor} might have slightly lower
performance compared to their regular peers. Our analysis of performance
proxies in Section \ref{ssec:proposal-proxies} will confirm this speculation (see Figure \ref{fig:proposal-theory-results}).
 
\subsection{Analysis of Performance Proxies}
\label{ssec:proposal-proxies}

\ps{Formulas and asymptotic analysis for diameter of regular arrangements}

\begin{figure}[h]
\centering
\captionsetup{justification=centering}
\begin{subfigure}{0.98 \columnwidth}
\centering
\includegraphics[width=0.95\columnwidth]{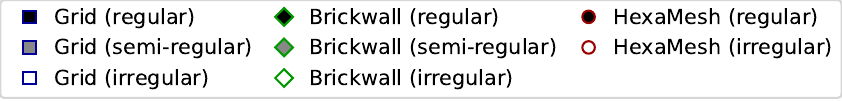}
\end{subfigure}
\begin{subfigure}{0.49 \columnwidth}
\centering
\includegraphics[width=1.0\columnwidth]{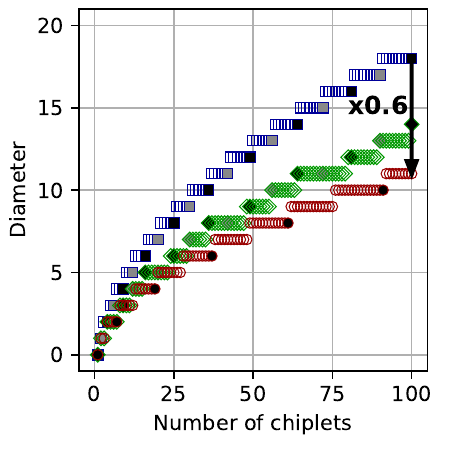}
\caption{Network Diameter.}
\label{fig:proposal-diameter}
\end{subfigure}
\begin{subfigure}{0.49 \columnwidth}
\centering
\includegraphics[width=1.0\columnwidth]{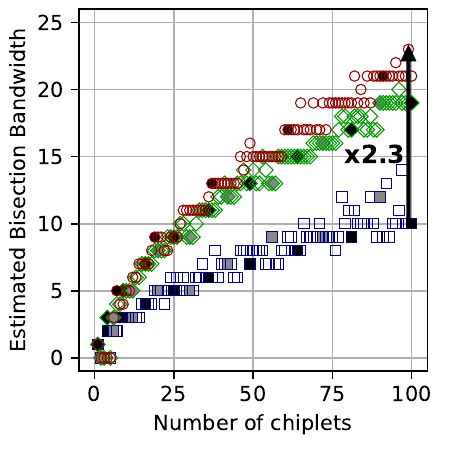}
\caption{Bisection Bandwidth.}
\label{fig:proposal-bandwidth}
\end{subfigure}
\caption{Performance proxies of chiplet arrangements.}
\label{fig:proposal-theory-results}
\vspace{-0.5em}
\end{figure}

\paragraph{Diameter}
The diameter for a regular \gls{g}, \gls{bw}, or \gls{cor} with $N$
chiplets can be computed as follows:
\begin{equation*} \label{eq:diam-grid}
	D_{G}\text(N) = 2\sqrt{N}-2
\end{equation*}
\begin{equation*} \label{eq:diam-brickwall}
	D_\text{BW}(N) = 2\sqrt{N}-2-\left \lfloor \sfrac{(\sqrt{N}-1)}{2} \right \rfloor
\end{equation*}
\begin{equation*} \label{eq:diam-circular}
	D_\text{HM}(N) = \sfrac{1}{3} \sqrt{12N-3} - 1
\end{equation*}
In Figure \ref{fig:proposal-diameter}, we compare the diameter of all three arrangements 
for chiplet counts from $1$ to $100$. 
The \gls{bw} has a significantly lower diameter than the \gls{g}, 
and the \gls{cor} further reduces the diameter.
We observe that regular and semi-regular \gls{g} and \gls{cor} provide the highest 
chiplet-count for a given diameter. 
For the \gls{bw}, regular and semi-regular arrangements do not seem to have 
advantages over their irregular peers. 
To analyze the asymptotic behavior of the diameter of regular arrangements, we compute
$\lim \limits_{N\to \infty}\frac{D_\text{BW}(N)}{D_\text{G}(N)} = \sfrac{3}{4}$ and 
$\lim \limits_{N\to \infty}\frac{D_\text{HM}(N)}{D_\text{G}(N)} = \sfrac{1}{\sqrt{3}}$. 
We conclude that asymptotically, the \gls{bw} reduces the diameter by 
$25\%$ and the \gls{cor} reduces the diameter by $42\%$ compared to the \gls{g}.

\ps{Formulas and asymptotic analysis for bisection bandwidth of regular arrangements}

\paragraph{Bisection Bandwidth}
The bisection bandwidth of a regular \gls{g}, \gls{bw}, or \gls{cor} with $N$
chiplets is computed as follows:
\begin{equation*} \label{eq:diam-grid}
	B_\text{G}(N) = \sqrt{N}
\end{equation*}
\begin{equation*} \label{eq:diam-brickwall}
	B_\text{BW}(N) = 2\sqrt{N}-1
\end{equation*}
\begin{equation*} \label{eq:diam-circular}
	B_\text{HM}(N) = \sfrac{2}{3} \sqrt{12N-3} - 1
\end{equation*}
Figure \ref{fig:proposal-bandwidth} compares the bisection bandwidth of all three 
arrangements for chiplet counts from $1$ to $100$. The bisection bandwidth of regular
arrangements is computed using the formulas above, that of semi-regular or irregular 
arrangements is estimated using METIS \cite{metis}.
The \gls{bw} comes with a significantly higher bisection bandwidth compared
to the \gls{g} and the \gls{cor} further improves upon the
\gls{bw}. 
To analyze the asymptotic behavior of the bisection bandwidth of regular arrangements, we 
compute
$\lim \limits_{N\to \infty}\frac{B_\text{BW}(N)}{B_\text{G}(N)} = 2$ and 
$\lim \limits_{N\to \infty}\frac{B_\text{HM}(N)}{B_\text{G}(N)} = \frac{4}{\sqrt{3}}$. 
Asymptotically, the \gls{bw} improves the bisection bandwidth 
by $100 \%$ and the \gls{cor} improves it by $130\%$
compared to the \gls{g}.

\section{A Model for \gls{d2d} Links}
\label{sec:model}

\ps{Section Intro}

The bisection bandwidth is an incomplete proxy for the global throughput, as it only considers the number of links, but not their bandwidth.
Since the \gls{bw} and \gls{cor} have more \gls{d2d} links per chiplet than the \gls{g}, 
the number of \gls{c4} bumps or micro-bumps per link and hence the per-link bandwidth is lower for them.
To estimate the link bandwidth for a given arrangement, we introduce our \gls{d2d} link model.

\subsection{Model Inputs}
\label{ssec:model-links}

\ps{Describe model inputs}

Table \ref{tab:model-inputs} lists the architectural parameters that our model needs as 
inputs to estimate the bandwidth of \gls{d2d} links.

\begin{table}[h]
\centering
\captionsetup{justification=centering}
\caption{Architectural Parameters Needed as Model Inputs.}
\begin{tabular}{ll}
\rowcolor{lightgray}
\hline
\textbf{Symbol} & \textbf{Description} \\
\hline
$A_{B}$ 	& \makecell[l]{Area (in mm$^2$) available for C4 bumps/micro-bumps 
			\\of one \gls{d2d} link}\\
\hline
$P_{B}$ 	& Pitch (in mm) of a C4 bump/micro-bump\\
\hline
$N_\text{ndw}$	& \makecell[l]{Number of non-data wires needed for a \gls{d2d} link 
			\\(e.g., wires for handshake, clock, etc.)}\\
\hline
$f$			& Frequency at which the \gls{d2d} links are operated\\
\hline
\end{tabular}
\label{tab:model-inputs}
\vspace{-1em}
\end{table}

\subsection{Link Bandwidth Estimation}
\label{ssec:model-link}

\ps{Explain Model for \gls{d2d} links}

We start by estimating the number of wires $N_w$ that can be built between the 
two chiplets. To compute $N_w$ we divide the area 
available for C4 bumps/micro-bumps by the squared pitch of said bumps. This estimate assumes a 
regular layout of bumps. A staggered layout would result in a slightly larger number of wires. \lbcom{Vague...}
To get the number of data wires $N_\text{dw}$ we subtract the number of non-data
wires $N_\text{ndw}$ from the number of wires $N_w$.
To estimate the link bandwidth $B$, we multiply the number of data wires $N_\text{dw}$
by the link frequency $f$.
\lbcom{I am a bit confused - f depends strongly on the length of the link... not really a constant as it depends on geometry...}
\picom{My initial idea was to estimate f based on the link length and required bit error rate, 
however, since my discussions with Matheus and Sina revealed that this relation is very complex,
I simply used f as an input parameter. It is still my goal to once have this model for f = function(BER, link-length) but
it seems very unrealistic to achieve this before the DAC deadline. Since in this paper, we only consider links between adjacent 
chiplets, the link-lengths are very similar, hence, this shortcoming of the model doesn't affect the results too much.}
\begin{equation*} \label{eq:model-Nw}
	N_{w} = \frac{A_{B}}{(P_{B})^2}
    \qquad \qquad
	N_\text{dw} = N_{w} - N_\text{ndw}
    \qquad \qquad
	B = N_\text{dw} \cdot f
\end{equation*}
In practice, the maximum operating frequency of \gls{d2d} links depends on the length of said links.
In this work, we only consider \gls{d2d} links between adjacent chiplets, whose lengths
are relatively short (below $4$mm in general, for $N\geq10$ chiplets even below $2$mm). Therefore,  we make the operating frequency
an input parameter rather than computing it based on the link's physical characteristics.
\begin{figure*}[t!]
\centering
\captionsetup{justification=centering}
\begin{subfigure}{1.0 \textwidth}
\centering
\includegraphics[width=0.9\textwidth]{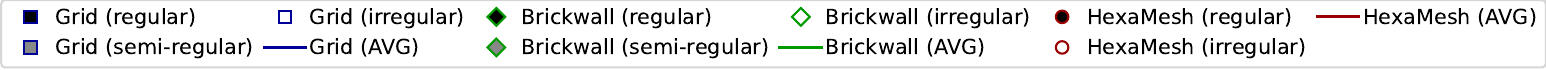}
\end{subfigure}
\\
\begin{subfigure}{0.24 \textwidth}
\centering
\includegraphics[width=1.0\textwidth]{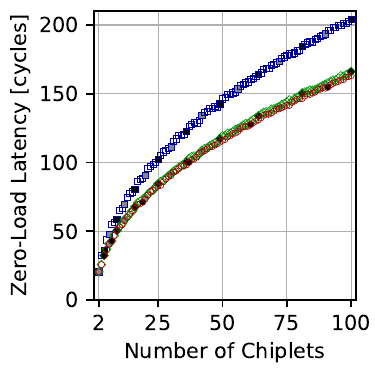}
\vspace{-2em}\caption{Latency.}\vspace{-0.5em}
\label{fig:evaluation-results-latency-abs}
\end{subfigure}
\begin{subfigure}{0.24 \textwidth}
\centering
\includegraphics[width=1.0\textwidth]{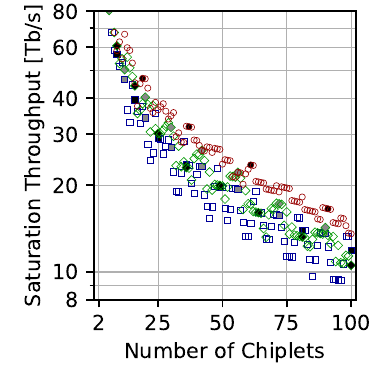}
\vspace{-2em}\caption{Throughput.}\vspace{-0.5em}
\label{fig:evaluation-results-throughput-abs}
\end{subfigure}
\begin{subfigure}{0.24 \textwidth}
\centering
\includegraphics[width=1.0\textwidth]{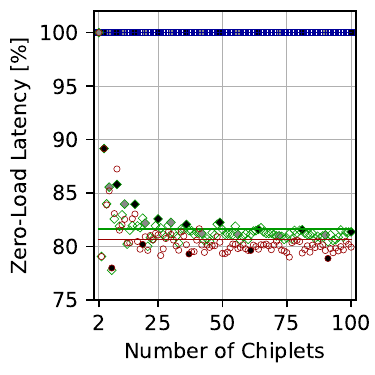}
\vspace{-2em}\caption{Normalized Latency.}\vspace{-0.5em}
\label{fig:evaluation-results-latency-rel}
\end{subfigure}
\begin{subfigure}{0.24 \textwidth}
\centering
\includegraphics[width=1.0\textwidth]{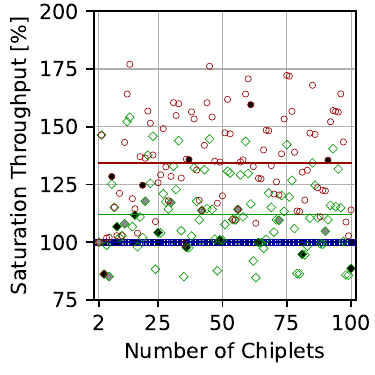}
\vspace{-2em}\caption{Normalized Throughput.}\vspace{-0.5em}
\label{fig:evaluation-results-throughput-rel}
\end{subfigure}
\caption{
Plots (a) and (b) show the zero-load latency saturation throughput of the \gls{g}, \gls{bw}, and \gls{cor}.
Plots (c) and (d) show zero-load latency and saturation throughput of the \gls{bw} and \gls{cor} relative to the \gls{g} (baseline).
\\\picom{Plot (d) looks messy, how can I improve it?}
}
\label{fig:evaluation-results}
\vspace{-1.5em}
\end{figure*}

\section{Evaluation}
\label{sec:evaluation}

\ps{Section Intro}

We leverage our model for \gls{d2d} links and network simulations in BookSim2 \cite{booksim}  
to compare the \gls{ici} performance of different arrangements of chiplets. 
Many parameters used in this section are based on
the UCIe protocol specifications \cite{ucie}.

\subsection{Cycle-Accurate Simulations using BookSim2}

\ps{Explain BookSim input parameters}
\label{ssec:evaluation-booksim}

We use the established, cycle-accurate BookSim2 \cite{booksim} network-on-chip simulator
to estimate the latency and throughput of different chiplet arrangements. 
The graph representation (see Section \ref{ssec:problem-proxies}) of a given arrangement 
is used as an input to BookSim2. We assume 
that each chiplet contains two endpoints and one local router. 
This router can route packets between the chiplet's PHYs or between a PHY and an endpoint.
We configure a link-latency of $27$ cycles which models the 
combined latency of the outgoing PHY, the \gls{d2d} link, and the incoming PHY
(UCIe \cite{ucie} states a \gls{phy} latency of $12$-$16$ UI).
Each router has a latency of $3$ cycles, $8$ virtual channels, and $8$ flit buffers.

\ps{How to extract BookSim results}

BookSim2 reports the average packet latency and the saturation throughput as the percentage 
of the full global bandwidth. The full global bandwidth is the maximum theoretical cumulative
throughput when all endpoints inject packets in the network at full rate; in our setting, it is
the product of the chiplet count, the number of endpoints per chiplet,
and the per-link bandwidth which we estimate using our model for \gls{d2d} links (see next paragraph).
We multiply the reported relative throughput by the
full global bandwidth of the corresponding arrangement to get the saturation throughput in Tb/s.

\subsection{Link Bandwidth Estimation using our Model}
\label{ssec:evaluation-model}

\ps{Explain \gls{d2d} link input parameters}

We use our model for \gls{d2d} links to estimate the per-link bandwidth in different 
arrangements of chiplets. To do this, we need to specify a set of architectural parameters.
We assume that the combined area of all chiplets is $A_\text{all} = 800$ mm$^2$ which is slightly
below the lithographic reticle limit. For an arrangement of $N$ chiplets, we compute
the chiplet area as $A_C = A_\text{all} / N$. We assume that any chiplet needs a fraction 
$p_p = 0.4$ of all \gls{c4} bumps for its power supply and that \gls{c4} bumps have a pitch of $P_B = 0.15$ mm.
Furthermore, we assume that $N_\text{ndw} = 12$ wires per link are needed for handshake and clock 
(UCIe \cite{ucie} uses $2$ clock-, $1$ valid- and $1$ track-wire per direction plus $4$ wires for the side band).
Finally, we assume that \gls{d2d} links are operated at $16$ GHz 
(UCIe \cite{ucie} can be operated at $16$ GHz to support its maximum data rate or $32$ GT/s).
The area $A_B$ available for bumps of a given \gls{d2d} link
is computed using the equations derived in Section \ref{ssec:proposal-shape} (except for
arrangements with $N \le 7$ chiplets which are hand-optimized).

\subsection{Discussion of Results}
\label{ssec:evaluation-results}

\ps{Discuss results (Figure \ref{fig:evaluation-results})}

Figures \ref{fig:evaluation-results-latency-abs} and 
\ref{fig:evaluation-results-throughput-abs} show the latency and throughput of the \gls{g}, 
\gls{bw}, and \gls{cor}, for chiplet counts from $2$ to $100$. Figures 
\ref{fig:evaluation-results-latency-rel} and \ref{fig:evaluation-results-throughput-rel}
show the latency and throughput of the \gls{bw} and \gls{cor} relative to
the \gls{g} (baseline). 
For $N \geq 10$ chiplets, both the \gls{bw} and the \gls{cor}
consistently reduce the latency by almost $20\%$ compared to the \gls{g}.
On average, the throughput is increased by $12\%$ if the \gls{bw} is used and by
$34\%$ if the \gls{cor} is used. We observe that the throughput relative to the
\gls{g} exhibits high fluctuations---this is mainly due to the inconsistent 
throughput of the \gls{g} (baseline). Another observation is 
that in practice (throughput), the \gls{bw} and \gls{cor} do not 
outperform the \gls{g} by as much as their theoretical superiority (bisection bandwidth) suggests.
The cause of this discrepancy is the fact that the \gls{bw} and \gls{cor} have more 
links per chiplet than the \gls{g} which results in fewer \gls{c4} bumps/micro-bumps per link and hence a lower
per-link bandwidth compared to the \gls{g}. This difference in bandwidth is accounted for in the 
simulations yielding the throughput but not in the theoretical analysis yielding the bisection bandwidth.
Nevertheless, we see that by using the \gls{cor}, we can significantly reduce the latency and 
significantly improve the throughput without adding any additional manufacturing complexity.
\section{Related Work}
\label{sec:related-work}

\ps{Discuss related work}

AMD \cite{amd-chiplets} shows how to use 2.5D integration to solve the economical challenges of technology scaling in production chips.
Since they use no more than eight compute-chiplets per chip, they can hand-optimize their chiplet arrangement.
In Tesla's Dojo training tile \cite{dojo} with $25$ chiplets where hand-optimizing the chiplet arrangement most likely is infeasible,
a 2D grid arrangement with a 2D mesh topology is used.
The mesh only connects adjacent chiplets which results in short links with high operating frequencies. 
Kite \cite{kite} is an \gls{ici} topology for 2D grid arrangements where non-adjacent chiplets are connected if the topological 
advantages of longer links outweigh their disadvantages due to a lower operating frequency.
By introducing HexaMesh, we achieve a low network diameter and a high 
bisection bandwidth while only connecting adjacent chiplets. This means that we can get rid of
the mesh's disadvantages (limited performance) while keeping its advantages (short, high-frequency links).
Coskun et al. \cite{placement-opt} introduce a cross-layer co-optimization approach for \gls{ici} design and chiplet arrangement.
For a set of existing topologies, they optimize the chiplet arrangement to maximize \gls{ici} performance and minimize 
manufacturing cost and operating temperature. In their approach, the chiplet arrangement depends on the topology while in our approach 
the topology depends on the chiplet arrangement (we connect adjacent chiplets). The advantage of our approach is that by only 
connecting adjacent chiplets, we minimize the length of \gls{d2d} links which maximizes their operating frequency.
There are many works in the 2.5D integration landscape that provide contributions orthogonal to ours.
Chiplet Actuary \cite{chiplet-actuary} for example provides a detailed cost model to 
analyze the economical benefits of disaggregation. This cost model could be applied together 
with our evaluation methodology to compare architectures both in terms of cost 
(Chiplet Actuary) and performance (our methodology).
Dehlaghi et al. \cite{usr-links} provide a detailed model to estimate the insertion loss and 
crosstalk of \gls{usr} \gls{d2d} links. Their work could be used to extend our model for 
\gls{d2d} links by adding predictions for the bit error rate in addition to our link bandwidth predictions.

\section{Conclusion}
\label{sec:conclusion}

\ps{Why chiplet shapes and arrangements matter}

2.5D integration is believed to be the solution to the economical challenges of CMOS technology scaling, but
it introduces a new challenge: Providing a high-performance inter-chiplet interconnect (ICI).
The fact that \gls{d2d} links need to be short to run at high frequencies strongly
limits the choice of \gls{ici} topologies. 

\ps{How we optimized chiplet shapes and arrangements}

In this work, we propose HexaMesh, an arrangement of chiplets that reduces the \gls{ici}'s network diameter
by $42\%$ while increasing its bisection bandwidth by $130\%$ compared to a grid arrangement. 
Furthermore, we introduce a model to estimate the bandwidth of \gls{d2d} links which is needed for
a fair comparison of designs with varying numbers of links per chiplet.
Our evaluations show that HexaMesh is not only superior to a grid arrangement in theory
but also in practice, as it reduces the latency by $19\%$ on average and improves the throughput by $34\%$ on average.
HexaMesh uses uniform and rectangular chiplets, which ensures that employing the HexaMesh arrangement 
does not increase the complexity of designing or manufacturing a chip.

\ifblind
\else
\section*{Acknowledgements}
\label{sec:acknowledgements}

This work was supported by the ETH Future Computing Laboratory (EFCL), financed by a donation from Huawei Technologies.
It also received funding from the European Research Council
\raisebox{-0.25em}{\includegraphics[height=1em]{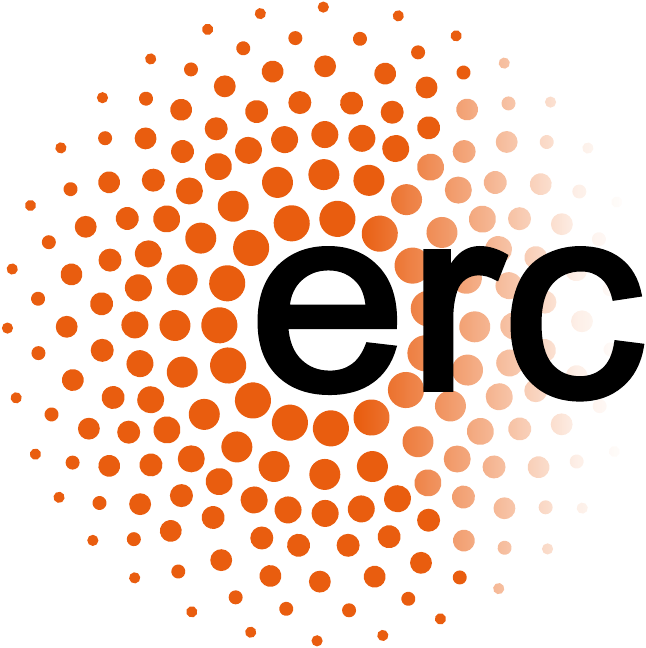}} (Project PSAP,
No.~101002047) and from the European Union's HE research 
and innovation programme under the grant agreement No.~101070141 (Project GLACIATION).
We thank Timo Schneider for help with computing infrastructure at SPCL.

\fi

\bibliographystyle{IEEEtran} 
\bibliography{bibliography}

\begin{thebibliography}{10}
\providecommand{\url}[1]{#1}
\csname url@samestyle\endcsname
\providecommand{\newblock}{\relax}
\providecommand{\bibinfo}[2]{#2}
\providecommand{\BIBentrySTDinterwordspacing}{\spaceskip=0pt\relax}
\providecommand{\BIBentryALTinterwordstretchfactor}{4}
\providecommand{\BIBentryALTinterwordspacing}{\spaceskip=\fontdimen2\font plus
\BIBentryALTinterwordstretchfactor\fontdimen3\font minus
  \fontdimen4\font\relax}
\providecommand{\BIBforeignlanguage}[2]{{%
\expandafter\ifx\csname l@#1\endcsname\relax
\typeout{** WARNING: IEEEtran.bst: No hyphenation pattern has been}%
\typeout{** loaded for the language `#1'. Using the pattern for}%
\typeout{** the default language instead.}%
\else
\language=\csname l@#1\endcsname
\fi
#2}}
\providecommand{\BIBdecl}{\relax}
\BIBdecl

\bibitem{design-cost-exp}
T.~Li, J.~Hou, J.~Yan, R.~Liu, H.~Yang, and Z.~Sun, ``Chiplet heterogeneous
  integration technology—status and challenges,'' \emph{Electronics}, vol.~9,
  no.~4, p. 670, 2020.

\bibitem{mooreslaw}
G.~E. Moore \emph{et~al.}, ``Cramming more components onto integrated
  circuits,'' 1965.

\bibitem{lookingglass2022}
S.~Mirabbasi, L.~C. Fujino, and K.~C. Smith, ``Through the looking glass---the
  2022 edition: Trends in solid-state circuits from {ISSCC},'' \emph{IEEE
  Solid-State Circuits Magazine}, vol.~14, no.~1, pp. 54--72, 2022.

\bibitem{amd-chiplets}
S.~Naffziger, N.~Beck, T.~Burd, K.~Lepak, G.~H. Loh, M.~Subramony, and
  S.~White, ``Pioneering chiplet technology and design for the {AMD} {EPYC}™
  and {RYZEN}™ processor families: Industrial product,'' in \emph{2021
  ACM/IEEE 48th Annual International Symposium on Computer Architecture
  (ISCA)}.\hskip 1em plus 0.5em minus 0.4em\relax IEEE, 2021, pp. 57--70.

\bibitem{bow}
``Bunch of wires (bow) phy specification,''
  \url{https://opencomputeproject.github.io/ODSA-BoW/bow_specification.html}.

\bibitem{ucie}
``{Universal Chiplet Interconnect Express (UCIe) Specification},''
  \url{https://www.uciexpress.org/specification}.

\bibitem{booksim}
N.~Jiang, D.~U. Becker, G.~Michelogiannakis, J.~Balfour, B.~Towles, D.~E. Shaw,
  J.~Kim, and W.~J. Dally, ``A detailed and flexible cycle-accurate
  network-on-chip simulator,'' in \emph{2013 IEEE international symposium on
  performance analysis of systems and software (ISPASS)}.\hskip 1em plus 0.5em
  minus 0.4em\relax IEEE, 2013.

\bibitem{intact}
P.~Vivet, E.~Guthmuller, Y.~Thonnart, G.~Pillonnet, C.~Fuguet, I.~Miro-Panades,
  G.~Moritz, J.~Durupt, C.~Bernard, D.~Varreau \emph{et~al.}, ``{IntAct}: A
  96-core processor with six chiplets {3D}-stacked on an active interposer with
  distributed interconnects and integrated power management,'' \emph{IEEE
  Journal of Solid-State Circuits}, vol.~56, no.~1, pp. 79--97, 2020.

\bibitem{usr-links}
B.~Dehlaghi, N.~Wary, and T.~C. Carusone, ``Ultra-short-reach interconnects for
  die-to-die links: Global bandwidth demands in microcosm,'' \emph{IEEE
  Solid-State Circuits Magazine}, vol.~11, no.~2, pp. 42--53, 2019.

\bibitem{stealth-dicing}
M.~Kumagai, N.~Uchiyama, E.~Ohmura, R.~Sugiura, K.~Atsumi, and K.~Fukumitsu,
  ``Advanced dicing technology for semiconductor wafer—stealth dicing,''
  \emph{IEEE Transactions on Semiconductor Manufacturing}, vol.~20, no.~3, pp.
  259--265, 2007.

\bibitem{plasma-dicing}
N.~Matsubara, R.~Windemuth, H.~Mitsuru, and H.~Atsushi, ``Plasma dicing
  technology,'' in \emph{2012 4th Electronic System-Integration Technology
  Conference}.\hskip 1em plus 0.5em minus 0.4em\relax IEEE, 2012, pp. 1--5.

\bibitem{graphs}
D.~B. West \emph{et~al.}, \emph{Introduction to graph theory}.\hskip 1em plus
  0.5em minus 0.4em\relax Prentice Hall Upper Saddle River, 2001, vol.~2.

\bibitem{metis}
G.~Karypis and V.~Kumar, ``Metis: A software package for partitioning
  unstructured graphs, partitioning meshes, and computing fill-reducing
  orderings of sparse matrices,'' 1997.

\bibitem{dojo}
E.~Talpes, D.~Williams, and D.~Das~Sarma, ``{DOJO: The Microarchitecture of
  Tesla's Exa-Scale Computer},'' in \emph{2022 IEEE Hot Chips 34 Symposium
  (HCS)}, IEEE.\hskip 1em plus 0.5em minus 0.4em\relax IEEE, 2022, pp. 1--28.

\bibitem{kite}
S.~Bharadwaj, J.~Yin, B.~Beckmann, and T.~Krishna, ``Kite: A family of
  heterogeneous interposer topologies enabled via accurate interconnect
  modeling,'' in \emph{2020 57th ACM/IEEE Design Automation Conference
  (DAC)}.\hskip 1em plus 0.5em minus 0.4em\relax IEEE, 2020, pp. 1--6.

\bibitem{placement-opt}
A.~Coskun, F.~Eris, A.~Joshi, A.~B. Kahng, Y.~Ma, A.~Narayan, and V.~Srinivas,
  ``Cross-layer co-optimization of network design and chiplet placement in
  {2.5-D} systems,'' \emph{IEEE Transactions on Computer-Aided Design of
  Integrated Circuits and Systems}, vol.~39, no.~12, pp. 5183--5196, 2020.

\bibitem{chiplet-actuary}
Y.~Feng and K.~Ma, ``Chiplet actuary: A quantitative cost model and
  multi-chiplet architecture exploration,'' \emph{arXiv preprint}, 2022.

\end{thebibliography}

\end{document}